\def\year{2021}\relax
\documentclass[letterpaper]{article} 
\usepackage{aaai21}  
\usepackage{times}  
\usepackage{helvet} 
\usepackage{courier}  
\usepackage[hyphens]{url}  
\usepackage{graphicx} 
\usepackage{natbib}  
\usepackage{caption} 
\nocopyright

\usepackage{booktabs}
\usepackage{tabu}
\usepackage{multirow}
\usepackage{amsmath}
\usepackage{xcolor}
\usepackage{soul}
\usepackage{makecell}
\usepackage{array}
\newcolumntype{L}[1]{>{\raggedright\let\newline\\\arraybackslash\hspace{0pt}}m{#1}}
\newcolumntype{C}[1]{>{\centering\let\newline\\\arraybackslash\hspace{0pt}}m{#1}}
\newcolumntype{R}[1]{>{\raggedleft\let\newline\\\arraybackslash\hspace{0pt}}m{#1}}

\newcommand{\xian}[1]{\textcolor{black}{#1}}
\newcommand{\addr}[1]{\textcolor{black}{#1}}
\newcommand{\revision}[1]{\textcolor{black}{#1}}
\newcommand{\rrv}[1]{\textcolor{black}{#1}}

\title{Characterizing User Susceptibility to COVID-19 Misinformation on Twitter}
\author {
    Xian Teng\textsuperscript{\rm 1},
    Yu-Ru Lin\textsuperscript{\rm 1},
    Wen-Ting Chung\textsuperscript{\rm 2},
    Ang Li\textsuperscript{\rm 1},
    Adriana Kovashka\textsuperscript{\rm 3} \\
}
\affiliations {
    \textsuperscript{\rm 1} School of Computing and Information, University of Pittsburgh \\
    \textsuperscript{\rm 2} Department of Psychology in Education, University of Pittsburgh \\
    \textsuperscript{\rm 3} Department of Computer Science, University of Pittsburgh \\
}

\begin{document}

\maketitle

\begin{abstract}
Though significant efforts such as removing false claims and promoting reliable sources have been increased to combat COVID-19 ``misinfodemic'', it remains an unsolved societal challenge if lacking a proper understanding of susceptible online users, i.e., those who are likely to be attracted by, believe and spread misinformation. This study attempts to answer {\it who} constitutes the population vulnerable to the online misinformation in the pandemic, and what are the robust features and short-term behavior signals that distinguish susceptible users from others. Using a 6-month longitudinal user panel on Twitter collected from a geopolitically diverse network-stratified samples in the US, we distinguish different types of users, ranging from social bots to humans with various level of engagement with COVID-related misinformation. We then identify users' online features and situational predictors that correlate with their susceptibility to COVID-19 misinformation. This work brings unique contributions: First, contrary to the prior studies on bot influence, our analysis shows that social bots' contribution to misinformation sharing was surprisingly low, and human-like users' misinformation behaviors exhibit heterogeneity and temporal variability. While the sharing of misinformation was highly concentrated, the risk of occasionally sharing misinformation for average users remained alarmingly high. Second, our findings highlight the political sensitivity activeness and responsiveness to emotionally-charged content among susceptible users. Third, we demonstrate a feasible solution to efficiently predict users' transient susceptibility solely based on their short-term news consumption and exposure from their networks. Our work has an implication in designing effective intervention mechanism to mitigate the misinformation dissipation.
\end{abstract}

\section{Introduction}

Even before the WHO declared COVID-19 a pandemic with accompanying ``massive infodemic'', the issues of COVID-19 misinformation and how it ramifies with the pandemic has been one of the foci for the science community, industry, public health professionals, and the general public. Abundant media and research reports have shown how the surge of misleading and false information spreading online triggered a wide range of socially harmful behaviors \cite{ognyanova2020misinformation,bridgman2020causes} that can jeopardize the trust of authority sources, effectiveness of the epidemic control measures and further impact all sectors of a society. While a significant research effort has been spent on cutting the supply of misinformation -- e.g., to fact-check or detect misinformation \cite{ball2020epic}, unless completely effective, cutting the supply alone will not solve the problem \cite{ball2020epic}. To intervene misinformation dissipation, there is a critical need in understanding users' (mis-)information behaviors.

Recent literature has shed light on the online misinformation phenomena. Several studies found that social bots are the main actors who spread low-credibility content, manipulated social media users' opinions and swayed democratic processes such as Brexit referendum and the US 2016 election \cite{shao2018spread,ferrara2016rise,bessi2016social}. Apart from the bot influence, Grinberg et. al. \cite{grinberg2019fake} discovered that a handful of ``supersharers'' and ``superconsumers'' accounted for nearly 80\% of fake news spread and consumption on Twitter during the past US presidential election. These studies together painted a picture of the key drivers in the fake news and misinformation campaigns that may influence political discourse and behaviors. In public health domain, the interlinking among misinformation and infectious diseases, such as the relationship with anti-vaccine attitude and other misconception, has been examined in other epidemics \cite{burki2019vaccine}. Since the COVID-19 outbreak, several studies have highlighted the ill effect of belief in false claims and its stemming on human's social and behavioral characteristics -- for example, people with particular cognitive tendency and sociopolitical identities may be more vulnerable to misinformation \cite{lazer2020state,grinberg2019fake,guess2020exposure}. Despite the rich literature, there is still no clear picture for how misinformation penetrates the {\it ordinary} users' information processing in the life during a pandemic. Rather than focusing on a particular role such as bots or super-users, we seek to understand what distinguish users who are actively engage with misinformation from those less or rarely.

Who -- bots or ordinary humans -- constitutes the population vulnerable to the online misinformation in the pandemic? To inform the design of timely intervention, can we answer this without knowing their cognitive tendencies and sociopolitical identity? Relying on observable social media traces, we focus on three research questions:


\begin{itemize}
\item[{\bf RQ1}] \xian{To what extent do ordinary humans, compared to bots and super users, contribute to the COVID-19 misinformation spread?} \addr{How do individuals engage with misinformation differently?}
\item[{\bf RQ2}] What are the robust features that distinguish susceptible users from those who have been resilient against the misinformation exposure?
\item[{\bf RQ3}] Can users' susceptibility to COVID-19 misinformation captured by time-varying, situational factors? To what extent we can identify users' risk of sharing misinformation purely depend on their short-term behavior traces?
\end{itemize}

\xian{To answer these RQs, we constructed a list of news sources flagged by journalists, fact-checkers and academia, and curated a longitudinal user panel on Twitter that were diverse in geography in United States. Our study period about COVID-19 misinformation covered the first 6 months of this pandemic. Our data also traced a user's timeline back to pre-pandemic period in order to understand the association between a user's past behaviors and misinformation engagement during this pandemic. Based on our data, we conducted a comprehensive analyses that compared user accounts by type (bots v.s. human-like users), by human-like users having different levels of misinformation engagement (RQ1) and behavioral characteristics (RQ2). Furthermore we exploited state-of-the-art deep learning models to perform real-time prediction purely based on short-term traces (RQ3). This work brings several unique contributions:}


\begin{itemize}
\item {\bf Analysis of misinformation engagement among the ordinary users.} Contrary to the prior studies on bot influence, our analysis shows that social bots' contribution to the misinformation sharing was surprisingly low, and majority of bots did not maliciously disseminate misinformation. On the other hand, {\it human-like} users' engagement with misinformation exhibit heterogeneity, where the sharing of misinformation was highly concentrated, but the risk of occasionally sharing and retweeting misinformation for average users remained alarmingly high.
\item {\bf Distinctive features correlated with users' susceptibility.} We identify features that distinguish users who are susceptible from those resilient to misinformation. \addr{Complementary to existing work, our results highlight the political sensitivity, activeness in tweeting and responsiveness to emotionally-charged content among susceptible users to COVID-19 misinformation.}


\item {\bf Predictive models for identifying future risk from short-term behavior traces.} \addr{Given that our analysis suggesting the transient nature of users' dynamic misinformation behaviors, we propose a feasible solution, using an interpretable deep learning approach, to identify the future risk of misinformation sharing solely based on users' weekly behavioral and exposure traces.}
\end{itemize}

\section{Related Work}

\subsubsection{{\bf COVID-19 Infodemic.}} \xian{To fight against the infodemic, researchers have been working to study the spread of misinformation on social media \cite{hameleers2020feeling,papakyriakopoulosspread,memon2020characterizing}, the prevalence, predictors, causes, and consequences of belief in COVID-19 false statements or/and conspiracy theories \cite{lazer2020state,roozenbeek2020susceptibility,uscinski2020people}.} 
\addr{It was found that citizens from US, UK, Netherlands and Germany experienced relatively high levels of mis-/disinformation -- 4.88/4.46 on a 7-point scale -- during the first stage of this pandemic \cite{hameleers2020feeling}. Although platforms' moderation practices were effective in reducing false claims, certain issues (e.g, timeliness, magnitude and moderation bias) were identified \cite{papakyriakopoulosspread}. Studies have also shown that a substantial proportion of respondents indicated that they believe COVID-19 conspiracy theories, for example, \citeauthor{lazer2020state} show that 7\%-22\% of US respondents believe 11 false claims about COVID-19 (e.g., coronavirus is created as a weapon in a Chinese lab, only older people get COVID-19), \citeauthor{roozenbeek2020susceptibility} show that 22–23\% respondents in UK, 26\% in Ireland, 33\% in Mexico and 37\% in Spain agreed that coronavirus was engineered in a laboratory in Wuhan.}
\xian{Such susceptibility might lead to undesirable consequences as studies have suggested that belief in false claims is negatively related to vaccine acceptance and self-reported compliance with guidelines \cite{roozenbeek2020susceptibility,lazer2020state,bridgman2020causes}.} \addr{Our work is different from prior work in several aspects: first, our study leverages large-scale social media traces instead of survey data to examine online users' susceptibility, second, our study is not limited to the several most popular COVID-19 conspiracy theories, instead we study the broad spectrum of online COVID-19 misinformation; third, our work considered a broad spectrum of online actors including bots, superusers and average users to understand to what extent and how (strategies) they contributed to misinformation spread.}

\subsubsection{{\bf The Fake-News Phenomena on Social Media.}} \xian{Fake news on social media has become a public concern. Existing work that are most related to ours include bots/trolls influence in promoting unreliable contents \cite{ferrara2016rise,ferrara2020types,badawy2019falls}, and predictors for users' susceptibility, e.g., such as demographics, socio-psychological attributes, political learning, news diets \cite{grinberg2019fake,allcott2017social}. It has been found that bots occupy a relatively high proportion (nearly 15\%) of Twitter accounts in 2016 US presidential election \cite{varol2017online,bessi2016social}, they are more active in posting and likely to mention influential accounts to promote low-quality content, therefore succeed to spread information into human population \cite{shao2018spread}. In addition to bots, studies have suggested that certain types of human users are more vulnerable to misinformation, i.e., people who are older \cite{grinberg2019fake}, less educated \cite{allcott2017social}, maintaining a more conservative news diet or/and supporting Trump \cite{grinberg2019fake,guess2020exposure}. Contrary to prior work concerning bot impact, this study finds that humans (not bots) have been heavily engaged with COVID-19 misinformation on Twitter. \rrv{This finding echos a recent study \cite{silva2020predicting}, indicating that 8.5\% were bots among those who tweeted COVID-19 misinformation, however, this study discarded all duplicated retweets which might fail to account for bot activities in resharing COVID-19 misinformation.}  Furthermore, we observed that humans exhibited heterogeneity in their misinformation behaviors in terms of volume, strategies and temporal dedication. Regarding susceptibility predictors, our results highlight the responsiveness of susceptible users to emotionally- and politically charged content.}




\subsubsection{{\bf Combating Mis-/disinformation.}} The computing community is engaging in the design of advanced methods to automatically combat mis-/disinformation on social media, including bot detection \cite{ferrara2016rise}, fake news detection \cite{shu2017fake} and cascade intervention \cite{sharma2020network}. \addr{Bot detection systems can be divided into three classes: techniques based on social network,
systems leveraging human intelligence,
and machine learning models based on discriminative features \cite{ferrara2016rise}.
However, the deployment of bot detection systems is undermined by the demand of significant amount of user-level information, and the challenge to account for bots' continuously changing behaviors \cite{ferrara2016rise}. Fake news detection techniques typically leverage information from news contents (texts, images and videos) and social contexts (user profiles, post responses, social networks) \cite{zhou2019fake}. A major issue is that fake news are constantly being produced in large scale following emerging and time-critical events, therefore easily spread to large audience without early warning or/and being fact-checked.} 
\revision{Cascade intervention leverages the understanding of diffusion dynamics to monitor a small set of influential nodes or intercept certain propagation paths to limit the spread of misinformation, yet it requires an overview of social network graphs.} \addr{Contrary to prior efforts, we propose a different task to battle misinformation, i.e., to proactively foresee a user's near-future susceptibility by solely using its short-term behavior traces. Our aim is to develop a situational and social-aware tool to catch up risk early before misinformation are further shared to larger population. For this purpose, we develop a model exploiting state-of-the-art deep learning techniques, which is able to capture the correlations of domains as well as dynamics of news ecosystem leveraging users' temporal cosharing/coexposure behaviors.}

\section{Data \& Study Design}
This section describes our data collection and processing, as well as crucial definitions and measurements.

\subsection{Data}\label{sec:data}

\subsubsection{Flagged News Sources.} \xian{We adopted a list of flagged news sources as proxy of misinformation following suggestions by prior work \cite{lazer2018science}, since using fact-checked articles might end up biasing towards a small set of ``popular'' or/and ``fact-checkable'' news stories.} \revision{Our flagged news source dictionary was constructed by referring to two datasets. (i) We first included the news sources curated by MediaBias/FactCheck (MBFC hereafter) under two categories -- conspiracy-pseudoscience and questionable sources (accessed on April 19, 2021), where the the conspiracy-pseudoscience category ``may publish unverifiable information that is not always supported by evidence'', and the questionable source ``exhibits one or more of the following -- extreme bias, consistent promotion of propaganda/conspiracies, poor or no sourcing to credible information, a complete lack of transparency and/or is fake news.'' We used the most updated version of MBFC news sources in April 2021 (prior to our paper submission),} \rrv{to included the most up-to-date information about websites that published COVID-19 misinformation during our study period.\footnote{\revision{We manually checked MBFC's analysis reports for flagged websites and found at least 143 websites were labeled by MBFC to be the sources spreading COVID-19 misinformation.}}} \revision{(ii) We also included several pre-existing lists of websites \cite{grinberg2019fake}, including three sets of publishers released by fact-checkers and journalists (i.e., published by Buzzfeed News, Politifact, and FactCheck.org), and two sets of problematic domains labelled by scholars \cite{guess2018selective,grinberg2019fake}. Following \citeauthor{grinberg2019fake}, we only included the domains if labeled as black, red or orange as these colors reflected annotators' stronger affirmation regarding a flawed editorial process (e.g., little regards for the truth, negligent or deceptive). In total, there were 1528 flagged news sources in our dictionary, and 527 of them appeared in our tweet data.}


\subsubsection{{User Panel Construction.}} \addr{To identify users contributed to COVID-19 information sharing,} we first referred to a public data of COVID-19 Twitter IDs between January 28 and April 24 2020 \cite{chen2020tracking}. 
To obtain a user panel that were diverse in geography and ideological context, we selected four cities (with a population larger than 250K people) spanning from liberal to conservative based on conservatism scores estimated by \cite{tausanovitch2014representation}, including San Francisco CA, New York City NY, Houston TX and Nashville TN.\footnote{Please note that the political diversity was not based on voting records. These four cities were chosen to achieve a trade-off between diversity and availability of user samples. We found that our user distribution across cities looks similar to the distribution obtained from the above public COVID-19 data crawled using keywords \cite{chen2020tracking}.} To understand users' social network properties, we randomly selected 100 seed users in each city, and snowball their friends to construct a social network. Specifically, the seeds were guaranteed to be active human accounts \revision{-- i.e., they were not detectable bots or organizational accounts (detected using tools discussed below) and they had posted at least two URLs prior to April 24, 2020.} We ensured that half of the seeds shared a proportion larger than 60\% of URLs from flagged sources while the other half shared less than 40\% URLs from those sources, so as to capture both resilient and susceptible users (formal definitions can be found in section \ref{sec:definition_studyperiod}). Starting from the seeds, we obtained up to 5000 users IDs they have been following (known as their ``friends'') as of June 2020. This step was run for two steps (therefore two-step snowball sampling) and resulted in more than 9.5 millions unique user IDs. We kept the accounts having at least one friend and one follower in the social network to make our panel in a reasonable size and facilitate our subsequent social network analyses. Then we crawled the most recent 3000 tweets posted by each user as of June 2020. In order to exclude accounts that were created after outbreak to be used to run COVID-19 related campaigns, we only kept users who registered their accounts before November 1 2019. We also removed inactive accounts who have not posted any tweets between December 1 2019 and June 1 2020. To guarantee we cover a user's complete activities after outbreak, we only kept users whose oldest tweet in our data was published prior to December 1 2019. Eventually, we ended up with 531,865 users and approximately 1.4 billions tweets.

\subsubsection{{Location Extraction.}} We extracted users' locations based on the location field within a user object. We matched the text against a list of cities, counties, states and city/state abbreviations in the US. For example, we extracted location ``Winslow, Arizona'' from the text ``on a corner in Winslow, Arizona,'' and ``NYC, NY'' from the text ``NYC, New York.'' In total, 287,784 users didn't disclose any city or state information, 244,081 users provided state information and 168,788 users provided city information.

\subsubsection{{COVID-19 Tweets.}} 
\xian{As our study is in the COVID-19 context, we particularly} identified COVID-19 tweets by searching for any of the COVID-19 related keywords constructed in \cite{chen2020tracking} within text, URLs, hashtags, mentions and screen names.


\subsubsection{{URL Expansion.}} Many users leveraged shortened links to share long URLs to maintain the maximum number of characters for posts. For the URLs included in COVID-19 tweets, we followed all redirects and obtained the final website domain. But for the URLs in COVID-19 irrelevant tweets, it would be impractical to expand them all within a reasonable time period. Therefore, we recovered a subset of shorted URLs from known URL shorteners to the original domains (e.g., ``youtu.be'' to ``youtube.com'', ``wapo.st'' to ``washingtonpost.com''), and merged with long URLs to be used in our subsequent analysis.

\subsection{Definitions and Measurements}
\label{sec:definition_studyperiod}

\subsubsection{Defining Study Period.} As described in data section \ref{sec:data}, our data traces a user's data before outbreak as well as during the first 6 months of this pandemic until early June 2020. Since the first human cases of COVID-19 were reported to be identified in Wuhan, China in December 2019, which triggered large-scale online discussion on social media ever since, we defined December 1 2019 to June 1 2020 as our COVID-19 misinformation study period $\mathbf{P}_{cov}$. We considered the time prior to December 1 2019 as pre-pandemic period $\mathbf{P}_{pre}$.

\subsubsection{Defining Susceptible/Resilient Users.}
\xian{We rely on the sharing activities of users to study misinformation engagement, defining a {\it share} as publishing a post (e.g., tweet, retweet, quote or reply) that included a URL redirecting to a page outside of Twitter.} A single post consisting of five external URLs would be counted as five shares. If the post were an original tweet, we would have five {\it original shares}. If one contained URL were from one of the flagged sources, we would call it a {\it problematic share}; We generally called URLs from unflagged sources as {\it other shares}. Accordingly, a {\it susceptible user} is defined as an account who had at least one problematic share, and a {\it resilient user} as who didn't share any URLs from flagged sources. Furthermore, we introduced a susceptibility adherence metric as level of engagement with content from flagged sources, measured by the number of problematic shares.
If restricting our scope to COVID-19 tweets, we have COVID-19 sharers, COVID-19 problematic shares, COVID-19 susceptible and resilient users. \xian{We will omit the term ``COVID-19'' in subsequent discussion if no special instruction. In summary, there are 265,859 sharers -- 88,160 are susceptible and 177,699 are resilient, who made 5,515,408 shares -- 560,788 are problematic shares and 4,954,620 other shares.}



\subsubsection{{Identifying User Types.}} We utilized a machine learning tool called {\it Humanizr} \cite{mccorriston2015organizations} to automatically identify organizational accounts (e.g., institutions, corporations etc). In total, 764 organizational accounts were located and excluded in our analysis. In order to further distinguish bot-like and human-like accounts, we used a bot detection tool system {\it Botometer} to assign a continuous score to each user \cite{varol2017online}. \rrv{To avoid time misalignment, our Botometer analysis was conducted immediately after we had collected Twitter user IDs between June 14 and July 9.}

\subsubsection{{Estimating News Medias Bias.}} We inferred a news media's political alignment by aggregating four distinct but overlapping datasets: two lists of news sources containing 500 and 224 sites from prior research \cite{bakshy2015exposure,grinberg2019fake}, a list of 1658 news medias labeled by Allsides.com, and a list of 312 news sites curated by MediaBias/FactCheck. Specifically, \citeauthor{grinberg2019fake} and \citeauthor{bakshy2015exposure} assessed a media's political alignment as the proportion of registered Republicans and Democrats who engaged (exposed or shared) with the source, thus a continuous score (-1 is most left, +1 right) is assigned to each website. Allside's media bias were obtained through a hybrid approach including editorial review, blind survey, third-party analysis and independent research. MediaBias/FactCheck's ratings were average scores by considering aspects including biased wording/headlines, factual/sourcing, story choices and political affiliation. The later two use categorical labels (left, left-center, center, right-center and right). In order to address the inconsistency of four lists, we employed an iterative imputation method ``missForest'' to impute missing values in four datasets \cite{stekhoven2012missforest}. In total, there were 2163 news medias in our analysis.

\subsubsection{{Identifying Right/Left-leaning Hashtags.}} Twitter users include hashtags in their user profile descriptions to connect with ideologically similar users and express support for movements/politicians. We extracted the most popular hashtags that were present in at least 100 users' descriptions. 
\xian{We identify a hashtag as ideologically relevant if it either supports or criticizes political leaders, their statements and campaigns (\#trump, \#maga, \#bluewave). To label a hashtag, we used Twitter’s search engine to retrieve relevant tweets, people, photos and videos, then examined the top ten results in each of those categories.} In total, there were 83 right-leaning and 69 left-leaning hashtags used in our analysis. \addr{Due to space limitation the list is provided in supporting file and will be made publicly available in our github repository.}


\section{Analysis \& Results}
This section reports our analysis and experimental results for three research questions.

\subsection{RQ1. To What Extent and How Differently Users Engage with Misinformation?} \label{rq1}

\subsubsection{4.1.1 Bots' Contribution.}\label{rq1a}
\begin{figure}[t]
  \centering
  \includegraphics[width=0.9\linewidth]{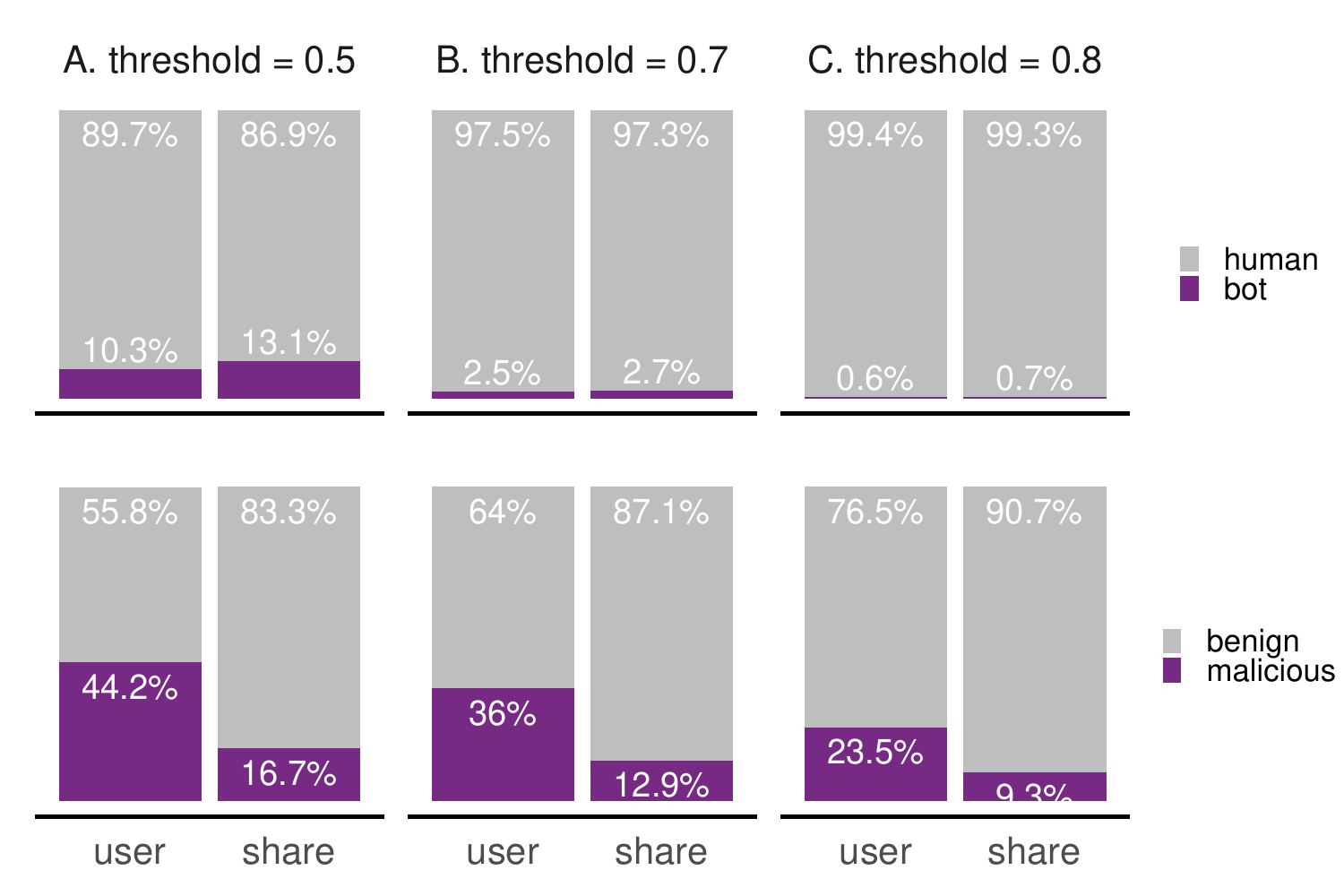}
  \caption{Prevalence of social bots estimated by varying bot score thresholds (A-C). Upper bars indicate separation of humans v.s. bots (left) and shares attributable to them (right). Lower bars indicate separation of benign v.s. malicious bots and shares attributable to them.}
  \label{fig:bots}
\end{figure}
First, we computed the proportion of social bots among all COVID-19 sharers. Following prior studies \cite{varol2017online,grinberg2019fake}, we labelled bots at thresholds 0.5 and 0.7: an account was considered to be a bot if its Botometer score were above 0.5 or 0.7. We added another threshold at 0.8 to achieve a posterior probability of 50\% (implying that with half chance an account having a score $\geq 0.8$ is a bot) for sensitivity test. Fig.~\ref{fig:bots} reveals that at threshold 0.5 approximately 10.3\% of misinformation sharers are bots, who are responsible for 13.1\% of problematic COVID-19 shares. The proportion of bots is even smaller -- 2.5\% and 0.6\% -- if setting threshold to 0.7 and 0.8, accordingly the percentage of associated shares drops to only 2.7\% and 0.7\%. Furthermore, we calculated the decomposition of different types of bots -- malicious v.s. benign. A bot was considered to be {\it malicious} if it had at least one problematic share, otherwise {\it benign}. We find that a large proportion of bots are actually benign -- the fraction is 55.8\% at threshold 0.5, 64\% at threshold 0.7 and nearly 76.5\% at threshold 0.8. By examining the decomposition of shares made by the two bot types, we find that benign bots are more active than malicious ones in sharing: at threshold 0.5, malicious bots only account for 16.7\% of all bot shares while benign bots are responsible for up to 83.8\% of all bot shares; The activity difference is even larger when increasing the threshold to 0.8, i.e., 9.3\% of shares (note that not all of them are problematic shares) are from malicious bots while nearly 90.7\% shares from benign bots (therefore not problematic). To summarize, we find that in the COVID-19 context the prevalence of social bots among problematic sharers is relatively low, and surprisingly malicious bots account for a very large fractions of presence and sharing activities. These findings in turn suggest that not bot but humans were mainly engaging with misinformation during our study period.


\subsubsection{4.1.2 Superusers' Contribution.}\label{rq1b}
\begin{figure}[t]
  \centering
  \includegraphics[width=0.9\linewidth]{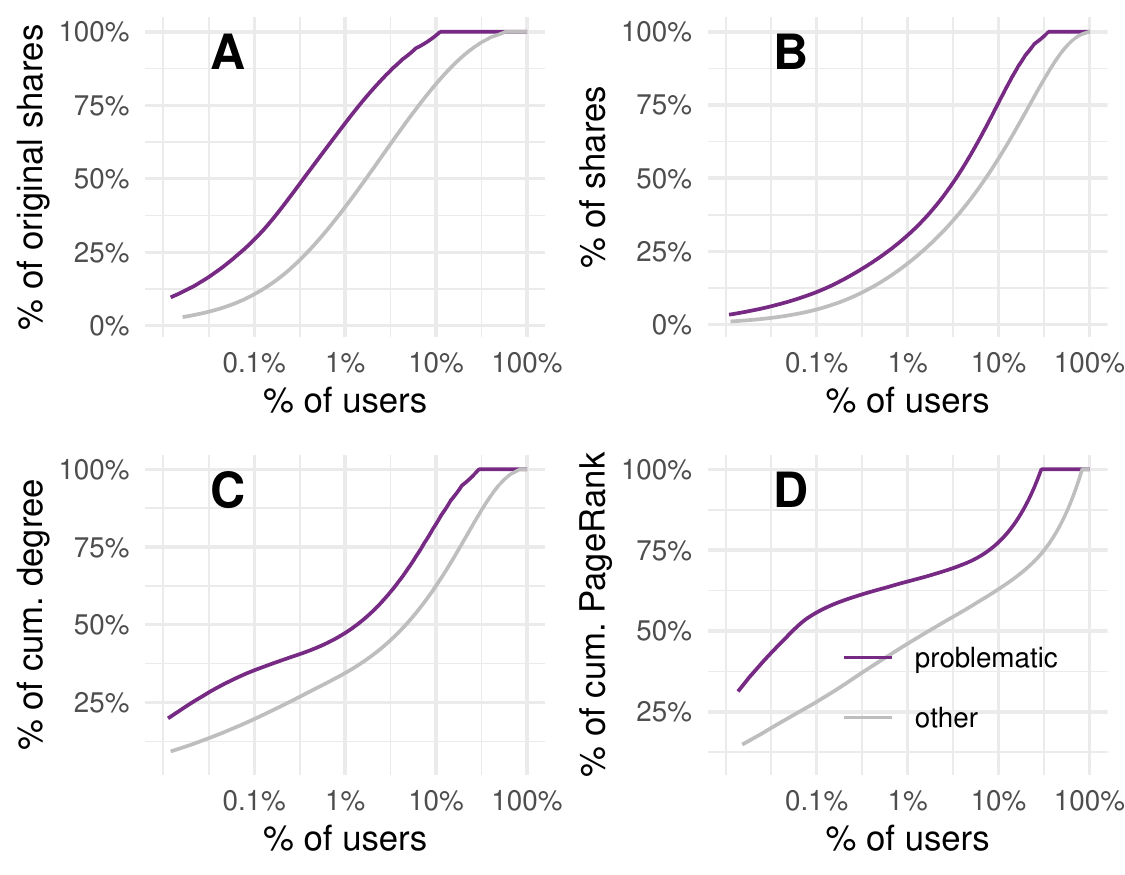}
  \caption{Concentration of misinformation engagement. Empirical cumulative distribution functions (ECDF) for number of shares in original tweets (A), number of shares in both original posts and retweets (B), degree and PageRank in retweet network (C,D). The x-axis represents the percentage of users responsible for a given percentage (y-axis) of cumulative values. Purple (gray) lines represent problematic (other) shares. }
  \label{fig:concentration}
\end{figure}
In subsequent analysis we excluded bot-like accounts using the lowest threshold 0.5 to guarantee that the remaining human sample do not contain bots. To test our hypothesis that the engagement with COVID-19 misinformation is heavily concentrated on a small fraction of core sharers, we computed four distinct metrics -- number of original shares, number of shares, the degree centrality and PageRank in retweet network. The retweet network was constructed as follows: if a user A retweeted B's post that was COVID-19 related and contained external URLs, we connected a directed edge from A to B. The former two metrics capture a user's level of sharing activity, while the latter two characterize a user's influence in information spread. It is revealed that merely 1\% of susceptible users account for almost 75\% of original shares (Fig.~\ref{fig:concentration}A), and 10\% of susceptible users account for 75\% of shares (Fig.~\ref{fig:concentration}B). Fig.~\ref{fig:concentration}C,D show that super 
influencers are present in COVID-19 retweet network -- if we consider degree and PageRank as indicators of ``wealth'', a 1\% of top-ranked susceptible users hold 50\% and 60\% of the cumulative wealth in the sample. The concentration phenomenon is more pronounced in problematic sharing (purple) compared to the sharing of other contents (gray).

\subsubsection{4.1.3 Heterogeneity of Humans' Misinformation Behaviors.}\label{sec:rq1c}
\begin{figure}[t]
  \centering
  \includegraphics[width=0.9\linewidth]{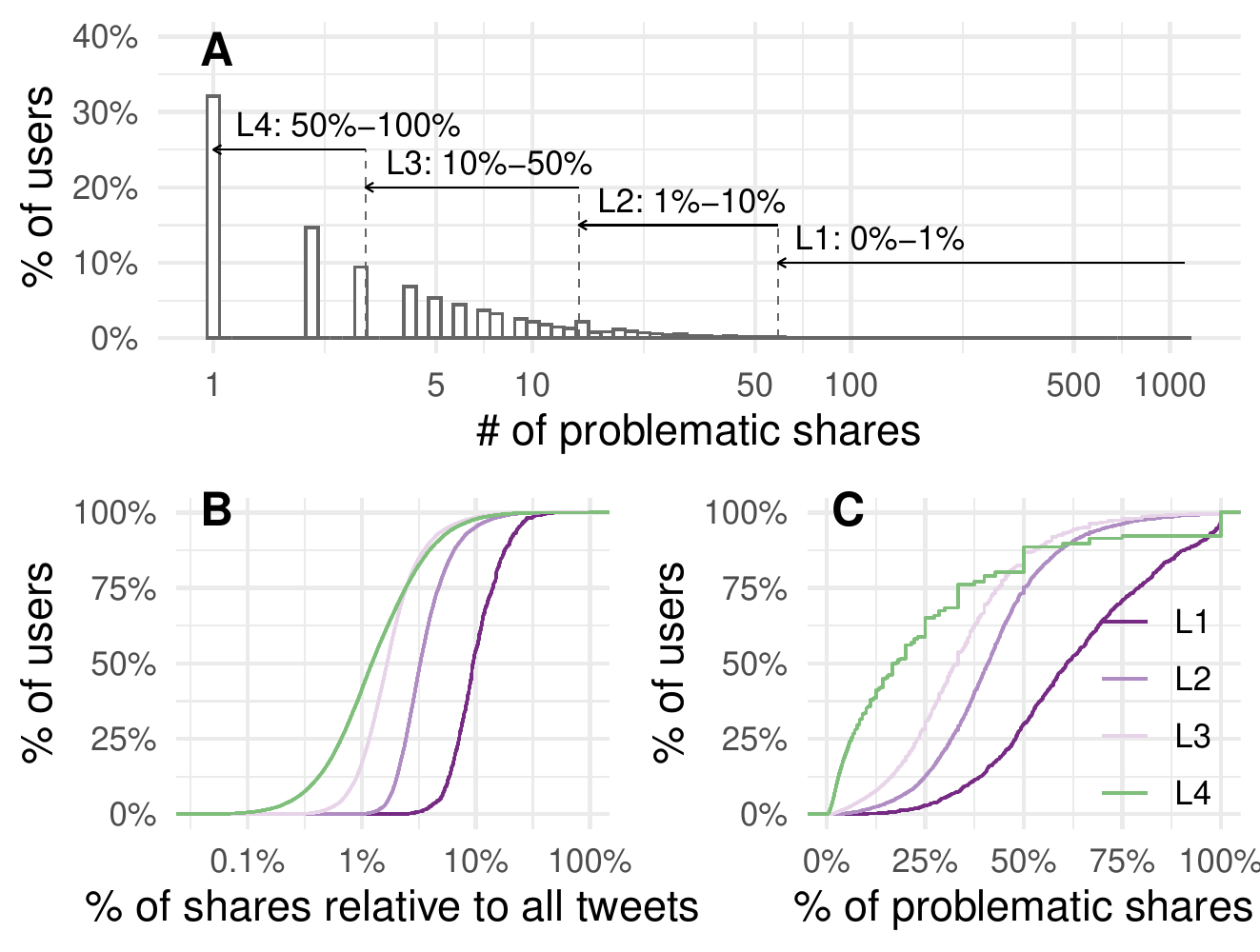}
  \caption{Dividing susceptible users into four levels. (A) Distribution of number of problematic shares and four levels L1-L4. (B-C) ECDF for the percentage of shares relative to COVID-19 tweets and the percentage of problematic shares relative to all COVID-19 shares. The y-axis indicates the percentage of users who have less than or equal to a certain value (x-axis).}
  \label{fig:levels}
\end{figure}
The preceding analysis suggests that users engaged with COVID-19 misinformation at different levels. We furthermore answered the question: how different are users who are strongly adherent to flagged sources, i.e., high-adherent users, compared to those who are weekly adherent? For this purpose, we divided susceptible human-like users into four levels: L1 included top 1\% users who had the largest number of problematic shares, L2 between 1\% and 10\%, L3 between 10\% and 50\%, and L4 between 50\% and 100\%. We found that nearly half of susceptible users only shared no more than two problematic URLs, raising the alarm that the risk of occasionally being exposed to or/and sharing misinformation is relatively high (Fig.~\ref{fig:levels}A). Fig.~\ref{fig:levels}B,C reveal that strong-adherent users (in contrast to weak-adherent users) have a stronger preference in using URLs when posting COVID-19 tweets, as the percentage of shares normalized by the number of COVID-19 tweets is relatively high (Fig.~\ref{fig:levels}C); They have a stronger interest in sharing information from flagged sources as the percentage of problematic shares normalized by number of shares is relatively high (Fig.~\ref{fig:levels}D). \xian{Specifically, we studied their differences in four aspects as follows.}

\begin{figure}[t]
  \centering
  \includegraphics[width=0.85\linewidth]{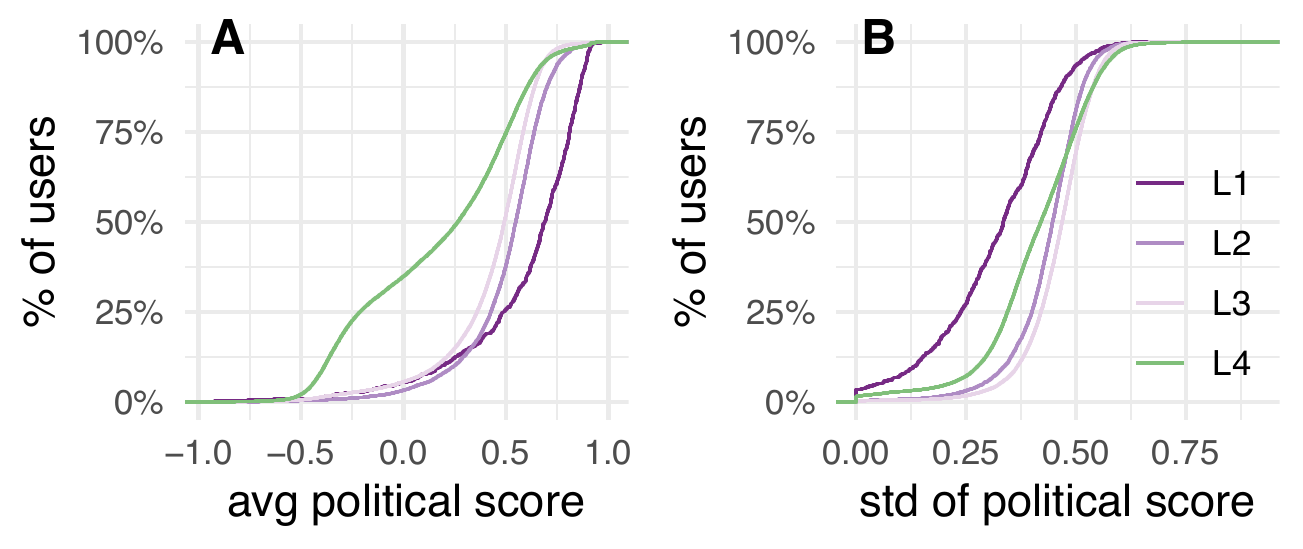}
  \caption{Consumption of news media for L1-L4 users. (A) Average political score of consumed news media. (B) Standard deviation of political scores.}
  \label{fig:media}
\end{figure}

\noindent {\bf (1) News source partisanship.} We computed the average and standard deviation of political scores of news sources shared by a user. A larger (smaller) score indicates that the user likes to share news from right-leaning (left-leaning) media sources, a larger (smaller) standard deviation implies that the user has broader (narrower) range of news preferences regarding politics. Fig.~\ref{fig:media}A,B show that high-adherent users are more likely to consume right-leaning news sources and hold narrower political news diet compared to individuals from L3-L4 groups.

\begin{figure}[t]
  \centering
  \includegraphics[width=0.85\linewidth]{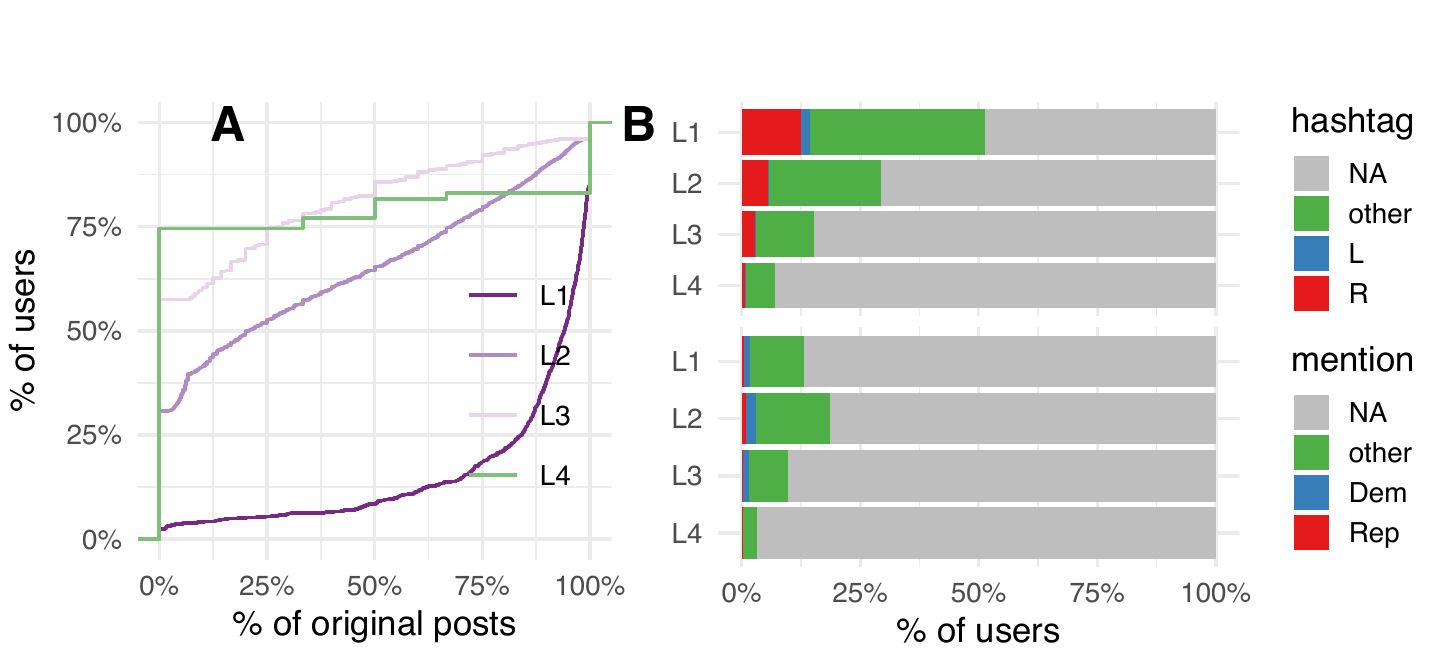}
  \caption{Tactics used in sharing COVID-19 misinformation for L1-L4 users. (A) ECDF for precent of original shares. (B) Percent of users who included hashtags and mentions in problematic shares.}
  \label{fig:tactics}
\end{figure}

\noindent {\bf (2) Sharing tactics.} \xian{We measured three possible tactics in sharing misinformation: (i) proactively introducing external misinformation into Twitter; (ii) exploiting hashtags -- particularly left/right-leaning hashtags -- to increase visibility of the shared misinformation; (iii) mentioning other accounts -- particularly Democratic or Republican leaders (senators and house representatives) -- while sharing misinformation. For (i), Fig.~\ref{fig:tactics}A reports the fraction of original shares relative to all problematic shares, suggesting that L1 users were more likely to introduce misinformation into Twitter, whereas more than half of L3-L4 individuals simply reshared the misinformation that had already existed. For (ii-iii), we calculated the percentage of individuals within each of the groups who had included certain hashtags or mentioned others in any of their problematic shares (Fig.~\ref{fig:tactics}B). We found that L1 users were prone to include hashtags -- especially right-leaning hashtags -- to increase visibility, whereas this association is not observed in the usage of left-leaning hashtags; Users rarely mentioned others while sharing misinformation, there is no significant association observed between mentions and misinformation engagement.}

\begin{figure}[t]
  \centering
  \includegraphics[width=\linewidth]{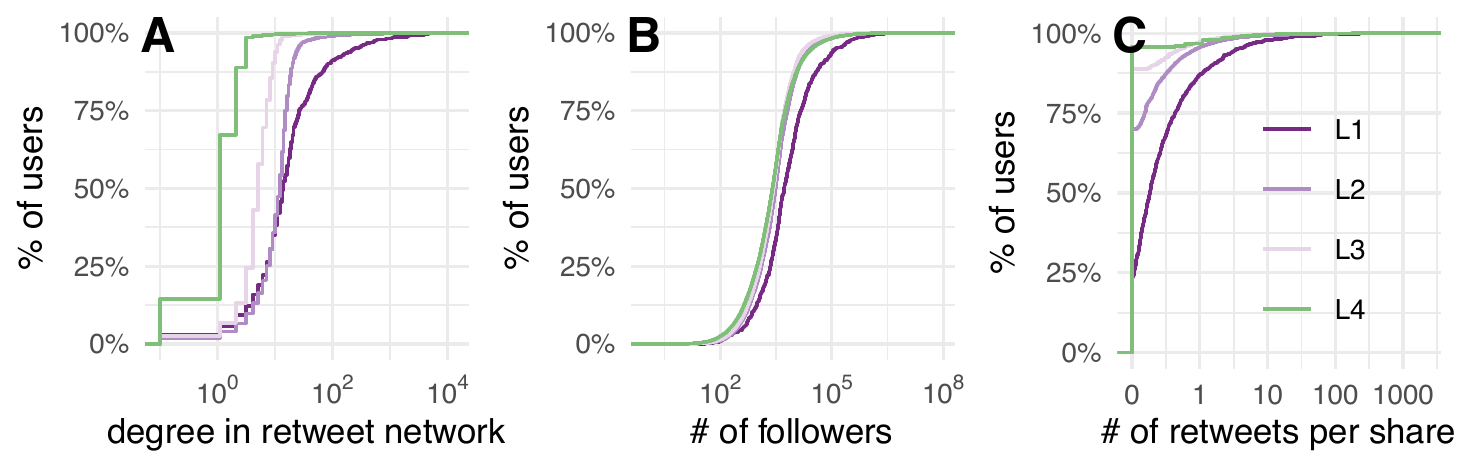}
  \caption{Social network properties for L1-L4 users. (A) ECDF of a user's degree in retweet network. (B) ECDF of an individual's followers counts. (C) ECDF for the number of retweets per problematic share.}
  \label{fig:net}
\end{figure}

\noindent {\bf (3) Social network characteristics.} \xian{We investigated to what extent high/weak-adherent users were located at important positions in retweet networks. Fig.~\ref{fig:net}A shows that L1 users have larger values of degree in retweet network (similar pattern is also observed for PageRank, we omit its plot to save space), suggesting that high-adherent users played a more successful role in spreading information into social network. One might suspect that perhaps they have more audience or/and were retweeted more often. So we compared the number of followers (Fig.~\ref{fig:net}B) and calculated the average number of times being retweeted per problematic share for users in L1-L4 groups (Fig.~\ref{fig:net}C). We found that high-adherent users didn't hold much structural advantages based on followers count, but they were more successful in misinformation diffusion process as L1 users were retweeted most often among all groups, approximately 80\% of L3-L4 users were not even retweeted by anyone in our data.}

\begin{figure}[t]
  \centering
  \includegraphics[width=\linewidth]{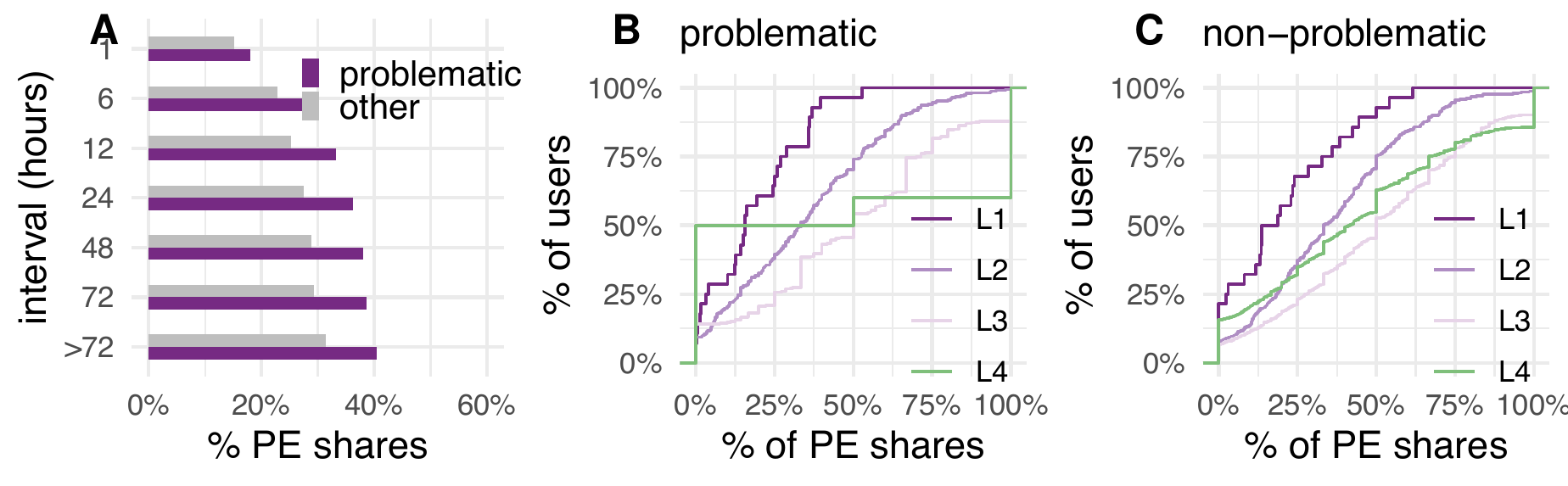}
  \caption{Exposure analysis from social context. (A) Fraction of shares that had a preceding exposure for both problematic and other-type information. (B) Fraction of problematic shares that had a preceding exposure for L1-L4 users. (C) Fraction of non-problematic shares that had a preceding exposure for L1-L4 users.}
  \label{fig:exposure}
\end{figure}

\noindent \rrv{\bf (4) Exposure of misinformation from social friends.} \rrv{
Before a user shares a URL at time $t$, they might have seen the same URL from their friends' tweets shared at an earlier time $t'$ ($t'<t$). We considered a potential exposure for an ego user to be any tweet shared by one of his/her friends, that includes the same URL he/she has shared. If multiple matches were found, we consider the most recent one $\Delta t=t-t'<\tau$ and break down the range of $\tau$ into different values ($\tau = 1, 6, 12, 24, 48, 72$, and $>$ 72 hours). If a share is successfully matched to a preceding exposure (PE), we call it a PE share, otherwise a non-PE share. We then consider the fraction of PE shares, i.e., the number of shares of type $c$ that has a PE / the number of all shares of type $c$, for $c\in\{$problematic, non-problematic$\}$.
Although a matched preceding exposure does not sufficiently imply a causal influence of sharing, this analysis allows us to capture the patterns of temporal ordering of exposure and sharing behaviors. Fig.~\ref{fig:exposure}A reports the percentage of PE shares for both types of shares (problematic and non-problematic) at different $\tau$'s. It shows that problematic shares from ego users, compared to non-problematic shares, are more likely to have a PE from friends, suggesting that misinformation might prone to spread through follow-followee relationship than reliable information \cite{vosoughi2018spread}. Fig.~\ref{fig:exposure}B,C shows the percentage of PE shares ($\tau\geq 0$) for problematic and non-problematic respectively, breaking down by L1-L4 users. Both figures suggest that average users (L2-L4) were more likely to reshare information from PE from their friends than L1 users.}

\noindent {\bf (5) Temporal variation of misinformation engagement.} \xian{Individuals' misinformation behaviors exhibited temporal variation, i..e, a user might share problematic URLs in certain weeks while not in others. We consider a user's status during a week as {\it SH} if he/she shared misinformation, and {\it NS} if didn't share any misinformation. Then we calculated the proportion of users in each of L1-L4 groups flip status across consecutive two weeks over a 3-month period between March and May. We find that on average L1 users have the highest tendency to maintain SH status (70\%$\pm$15\%), while L4 users have the highest tendency to maintain NS status (80\%$\pm$4\%). In contrast, L2/L3 users were more likely than L1/L4 users to change status, for example, 14\%-17\% of them flipped their status while this number is only 5\% for L1 and 10\% for L4.}

\noindent \xian{Overall, we find that users who heavily engaged with misinformation were more likely to share conservative websites, more active in introducing misinformation into the platform rather than simply forwarding it, and more strategic in promoting its spread. \revision{Average users, in contrast to those who heavily engaged with misinformation, have a tendency to reshare information after their friends' sharing.} The risk of occasionally sharing misinformation for average users remained high.} 


\begin{table}[p]
\centering \fontsize{8}{9}\selectfont \renewcommand\cellalign{lc}
\begin{tabular}{m{0.14\textwidth}m{0.07\textwidth}m{0.07\textwidth}m{0.07\textwidth}}
\makecell{\bf Features} & \makecell[c]{\bf ALL} & \makecell[c]{\bf PSU} & \makecell[c]{\bf PRU} \\
\midrule
\makecell[l]{Previously\\susceptible}& \makecell[c]{1.1***\\(1.0,1.1)}& \makecell[c]{}& \makecell[c]{}\\
\midrule
\makecell[l]{Swing county}& \makecell[c]{0.0\\(-0.0,0.1)}& \makecell[c]{0.1\\(0.0,0.1)}& \makecell[c]{-0.0\\(-0.1,0.0)}\\
\makecell[l]{Republican county}& \makecell[c]{0.2***\\(0.1,0.2)}& \makecell[c]{0.2***\\(0.2,0.2)}& \makecell[c]{0.2**\\(0.1,0.2)}\\
\midrule
\makecell[l]{\% leadership\\interactions - Rep}& \makecell[c]{10.0***\\(9.3,10.8)}& \makecell[c]{9.5***\\(8.5,10.5)}& \makecell[c]{9.1***\\(7.8,10.4)}\\
\makecell[l]{\% leadership\\interactions - Dem}& \makecell[c]{4.0***\\(3.5,4.5)}& \makecell[c]{4.7***\\(4.0,5.4)}& \makecell[c]{3.5***\\(2.8,4.2)}\\
\makecell[l]{\% retweet - Rep}& \makecell[c]{0.4**\\(0.3,0.5)}& \makecell[c]{0.3**\\(0.2,0.5)}& \makecell[c]{0.3\\(0.1,0.5)}\\
\makecell[l]{\% retweet - Dem}& \makecell[c]{-0.8***\\(-0.9,-0.7)}& \makecell[c]{-0.8***\\(-0.9,-0.6)}& \makecell[c]{-0.7***\\(-0.9,-0.6)}\\
\makecell[l]{\% reply - Rep}& \makecell[c]{-0.5**\\(-0.7,-0.3)}& \makecell[c]{-0.2\\(-0.5,0.1)}& \makecell[c]{-1.2***\\(-1.6,-0.9)}\\
\makecell[l]{\% reply - Dem}& \makecell[c]{0.6***\\(0.5,0.8)}& \makecell[c]{0.4*\\(0.2,0.6)}& \makecell[c]{0.6**\\(0.4,0.9)}\\
\makecell[l]{\% mention - Rep}& \makecell[c]{-0.4***\\(-0.5,-0.3)}& \makecell[c]{-0.4**\\(-0.6,-0.3)}& \makecell[c]{-0.2\\(-0.4,-0.1)}\\
\makecell[l]{\% mention - Dem}& \makecell[c]{-0.3**\\(-0.4,-0.2)}& \makecell[c]{-0.2\\(-0.3,-0.0)}& \makecell[c]{-0.4**\\(-0.6,-0.3)}\\
\makecell[l]{Sent. reply - Dem}& \makecell[c]{-0.2**\\(-0.2,-0.1)}& \makecell[c]{-0.1\\(-0.2,-0.0)}& \makecell[c]{-0.2*\\(-0.3,-0.1)}\\
\makecell[l]{Sent. reply - Rep}& \makecell[c]{0.1*\\(0.1,0.2)}& \makecell[c]{0.1\\(-0.0,0.1)}& \makecell[c]{0.2\\(0.1,0.4)}\\
\makecell[l]{Sent. mention - Dem}& \makecell[c]{-0.2***\\(-0.3,-0.2)}& \makecell[c]{-0.2**\\(-0.3,-0.1)}& \makecell[c]{-0.2**\\(-0.3,-0.1)}\\
\makecell[l]{Sent. mention - Rep}& \makecell[c]{0.1\\(0.0,0.2)}& \makecell[c]{0.1\\(0.0,0.2)}& \makecell[c]{0.0\\(-0.1,0.1)}\\
\makecell[l]{Right hashtags\\in description}& \makecell[c]{1.0***\\(0.9,1.1)}& \makecell[c]{0.7***\\(0.6,0.8)}& \makecell[c]{2.8***\\(2.6,3.1)}\\
\makecell[l]{Left hashtags\\in description}& \makecell[c]{0.4***\\(0.3,0.4)}& \makecell[c]{0.3***\\(0.3,0.4)}& \makecell[c]{0.5***\\(0.5,0.6)}\\
\midrule
\makecell[l]{\# tweets per day}& \makecell[c]{5.9***\\(5.4,6.4)}& \makecell[c]{5.7***\\(5.0,6.4)}& \makecell[c]{6.3***\\(5.5,7.1)}\\
\makecell[l]{\% interval $<$ 1 hour}& \makecell[c]{1.7***\\(1.6,1.8)}& \makecell[c]{1.3***\\(1.1,1.4)}& \makecell[c]{2.7***\\(2.5,2.9)}\\
\makecell[l]{\% interval $>$ 1 day}& \makecell[c]{-1.7***\\(-1.9,-1.5)}& \makecell[c]{-2.4***\\(-2.6,-2.2)}& \makecell[c]{-0.6*\\(-0.9,-0.3)}\\
\makecell[l]{\% tweets incl.\\hashtags}& \makecell[c]{-1.2***\\(-1.3,-1.2)}& \makecell[c]{-1.3***\\(-1.4,-1.2)}& \makecell[c]{-1.0***\\(-1.2,-0.8)}\\
\makecell[l]{\% tweets incl.\\mentions}& \makecell[c]{0.4***\\(0.3,0.5)}& \makecell[c]{0.2\\(0.1,0.3)}& \makecell[c]{0.7***\\(0.5,0.9)}\\
\makecell[l]{\% tweets incl. urls}& \makecell[c]{0.2**\\(0.1,0.3)}& \makecell[c]{0.4***\\(0.3,0.5)}& \makecell[c]{-0.1\\(-0.2,0.1)}\\
\makecell[l]{Followers/followees\\ratio (logged)}& \makecell[c]{-0.1**\\(-0.1,-0.0)}& \makecell[c]{-0.0\\(-0.0,0.0)}& \makecell[c]{-0.2***\\(-0.2,-0.1)}\\
\makecell[l]{Age (months)}& \makecell[c]{-0.2***\\(-0.3,-0.2)}& \makecell[c]{-0.3***\\(-0.4,-0.2)}& \makecell[c]{-0.0\\(-0.2,0.1)}\\
\makecell[l]{Account verified}& \makecell[c]{-0.5***\\(-0.5,-0.4)}& \makecell[c]{-0.5***\\(-0.6,-0.5)}& \makecell[c]{-0.3***\\(-0.4,-0.3)}\\
\midrule
\makecell[l]{Positive emotion}& \makecell[c]{-0.8***\\(-0.9,-0.6)}& \makecell[c]{-1.2***\\(-1.4,-1.0)}& \makecell[c]{-0.1\\(-0.4,0.2)}\\
\makecell[l]{Negative emotion}& \makecell[c]{3.2***\\(3.0,3.4)}& \makecell[c]{2.9***\\(2.7,3.2)}& \makecell[c]{3.3***\\(2.9,3.6)}\\
\midrule
\makecell[l]{Avg. political score}& \makecell[c]{2.1***\\(2.1,2.2)}& \makecell[c]{2.6***\\(2.5,2.7)}& \makecell[c]{0.5***\\(0.4,0.6)}\\
\midrule
N & \makecell[c]{44925} & \makecell[c]{25067} & \makecell[c]{19858} \\
AIC & \makecell[c]{36866} & \makecell[c]{23486} & \makecell[c]{12848} \\
Null deviance & \makecell[c]{55880} & \makecell[c]{34584} & \makecell[c]{15313} \\
Residual deviance & \makecell[c]{36802} & \makecell[c]{23424} & \makecell[c]{12786} \\
\midrule
& \multicolumn{3}{r}{*p$<$0.1, **p$<$0.05, ***p$<$0.001}
\end{tabular}
\caption{\revision{Coefficients (std errors) from classifying COVID susceptible v.s. resilient users among (i) {\bf All} users, (ii) {\bf PSU}: the subgroup who used to be susceptible prior to this pandemic, and (iii) {\bf PRU}: the subgroup used to be resilient.}}
\label{tab:features}
\end{table}

\subsection{RQ2. Discriminative Features}\label{rq2}

\xian{RQ1 overviews to what extent as well as how various actors engaged in spreading misinformation, which reveals the heterogeneity among the susceptible population. In RQ2, we particularly zoom into the major subset of ordinary susceptible users, investigating the contexts and behaviors -- in contrast against resilient users -- to better understand the predictors that correlate with being susceptible to misinformation.}

\subsubsection{\bf 4.2.1 Comparison Groups.}
To obtain a panel of ordinary individuals we first excluded organizational accounts, bots (at threshold 0.5), and superusers (L1 group), we kept accounts with city information to further guarantee they are genuine users. We then further cleaned our samples by removing IQR outliers (in terms of friends count, followers count and number of tweets posted per day). We limited our samples to those who have posted at least 10 tweets in $\mathbf{P}_{pre}$, 10 tweets in $\mathbf{P}_{cov}$ and 4 COVID-19 shares. \xian{We examined each panelist's misinformation behaviors in $\mathbf{P}_{pre}$ and $\mathbf{P}_{cov}$ to assign two types of labels accordingly. There are 63,371 users in our final penal -- 15,830 of them are {\it COVID-19 susceptible} among which 77\% were {\it pre-pandemic susceptible}, and 47,540 {\it COVID-19 resilient} among which 65\% were {\it pre-pandemic resilient}. For brevity, we denote COVID-19 susceptible/resilient users as CSUs/CRUs, pre-pandemic susceptible/resilient users as PSUs/PRUs below.}

\subsubsection{\bf 4.2.2 Features.} \xian{We examined a series of contextual variables.} \xian{{\bf (D1) Pre-pandemic susceptibility} describes whether a CSU/CRU used to be susceptible or not. {\bf (D2) Geopolitical environment} characterizes the political climate of the county a user lives in using returns of US 2000-2016 presidential elections, e.g., a swing county or not, a Democratic or Republican county. \revision{{\bf (D3) Political sensitivity} contains a user's tendency to interact with political leaders of Republican or Democratic party, the sentiment expressed in the interactions with them, as well as the variables indicating whether a user  used left/right-leaning hashtags in profile descriptions. For leadership interaction, we examined the overall proportion of interactions with leaders from one party among all interactions with others, as well as the percentage of different types -- retweet, mention and reply -- among leadership interactions; for sentiment, we calculated the median score (between -1 and +1) along with each type of leadership interactions.} {\bf (D4) Activeness} characterizes a user's activity pattern from distinct aspects such as tweeting intensity (tweeting rate and intervals), platform engagement (the proportion of using hashtags, URLs and mentions in tweets, ratio of followers to followees) and credibility (account age, verified or not). {\bf (D5) Emotions} are based on LIWC lexicons by computing the user-level proportion of tweets containing tokens from each lexicon category. {\bf (D6) News diets} captures the average political alignment score for news sources consumed by a user. {\bf (D7) Social exposure} include the fractions of right (political score $\leq$ -0.2), left (political score $\geq$ 0.2) and neutral friends in social network.}


\subsubsection{\bf 4.2.3 Findings and Insights.} \xian{Table~\ref{tab:features} shows logistic regression analyses from three models: column ``ALL'' corresponds to the model that includes pre-pandemic susceptibility as one of the predictors, while column ``PSU'' and ``PRU'' correspond to the one that classifies CSUs/CRUs among the subgroup of users who used to be susceptible (PSU) and resilient (PRU) prior to this pandemic, respectively. It reveals that the misinformation engagement was positively associated with being previously susceptible (D1), coming from Republican counties (D2) and having conservative news diets (D6). \revision{Interestingly, in overall CSUs are more likely than CRUs to interact with political leaders from both sides \rrv{($p<10^{-3}$)} -- Democratic and Republican, as well as to insert partisan hashtags \rrv{($p<10^{-3}$)} -- both left and right -- in their profile descriptions \rrv{(D3)}. When we further examined interaction types, we found that CSUs have a stronger tendency to reply to rather than to retweet the posts of Democratic leaders. In terms of sentiment, we observed that CSUs like to express greater negative sentiment towards Democratic leaders, particularly when they mention or reply to a Democratic leader.} Regarding activeness (D4), CSUs are more likely than CRUs to tweet frequently with shorter intervals, have accounts being unverified, but less likely to include general hashtags in tweets. CSUs have a tendency than CRUs to tweet with words showing negative emotions (D5). In terms of social exposures (D7), we found that CSUs have significantly more right-leaning friends (K-S $D=0.56, p<10^{-3}$), fewer left-leaning (K-S $D=0.54, p<10^{-3}$) and neutral friends (K-S $D=0.19, p<10^{-3}$) compared to CRUs.}

The positive association of sharing misinformation with conservative news diets, geopolitical environment and social exposures can be linked to prior literature, as existing work have shown that conservatives are more likely than liberals to 
believe conspiracy theories \cite{enders2019informational}, tolerate misinformation from politicians \cite{roets2019there}, distrust mainstream news media outlets and heavily consume hyper-partisan right-wing media outlets which lack fact-checking or editorial norms \cite{marwick2017media}. Besides, our findings regarding COVID-19 susceptible users' overall political sensitivity and negative emotions are in alignment with prior evidence that individuals who likely to engage with flagged sources are also highly engaged with political news \cite{grinberg2019fake}, and conservatives are more attracted than liberals by negativity and respond to various controversial issues with negative emotions \cite{inbar2009conservatives}. 
\revision{Our study highlights the patterns of susceptible users interacting with political figures from the opposing party: susceptible users are more likely to reply Democratic leaders instead of retweeting them (avoid endorsement or propagation). \rrv{Besides,} they choose to write negative words by replying to and directly mentioning Democratic leaders to express their opinions. These findings align with prior studies about adversarial interactions against candidates for the U.S. House of Representatives \cite{hua2020towards,hua2020characterizing}, which capture the cases of Twitter \rrv{users' hostility} using misinformation to attack and \rrv{to undermine} the legitimacy of candidates from opposing party, and those highly adversarial users express partisanship bias in their profiles.}
Furthermore, our results regarding activeness features can be explained by the difference in digital strategies of left/right-wing actors. Studies have shown that the left tend to directly rely on platforms to distribute messages, such as ``hashtag activism'' \cite{freelon2020false}, by contrast the right believe that the ``Big Tech'' platforms are biased against them and they manipulate those platforms to amply their messaging to larger audience, e.g., gaming Twitter's trending topics feature, using fake accounts, leveraging partisan hashtags to find ideologically similar users \cite{marwick2017media}. Therefore, CSUs exhibited higher tweeting rate, shorter intervals, accounts being unverified, less general hashtags in tweets but more partisan hashtags in profiles, compared to CRUs.

\subsection{{RQ3. Situational Prediction of Susceptibility}}
\revision{Our results in RQ1 suggest that a user's susceptible state keep changing over time (section 4.1.3 (5)), as well as possibly triggered by prior exposure of (mis)information from the his/her the continuously evolving social environments (section 4.1.3 (4)). Therefore, in RQ3 we ask the question: can we develop an situational, social-aware model to predict a user's near-future susceptibility based on his/her recent sharing activities and exposure history from friendships? This task is particularly crucial in practical settings where we need to distribute attention resources to at-risk users and catch up misinformation early before they are further shared to a larger population.}

\subsubsection{4.3.1 Prediction Task and Model.} 
\xian{In our prediction task, the target variable takes value 1 if a user, at a certain time point, shares misinformation (i.e., any URLs from flagged news sources) in the subsequent 7 days, 0 otherwise. The inputs include a sequence of website domains shared by the user during the past 2 weeks, called {\it share sequence}. Furthermore, to capture the possible trigger from social context exposure, we also generate another sequence of URLs shared by its 1-hop friends, called {\it exposure sequence}. We developed an interpretable deep learning model exploiting Transformer encoder to process sequential inputs \cite{vaswani2017attention}, where the intrinsic interpretability comes from the design of a linear format, i.e., each domain's contribution to the final prediction could be captured by a corresponding coefficient (illustrated in Fig.~\ref{fig:cases}). For comparison we implemented a logistic regression (LR) classifier that takes the counts of news sources encountered in the past as feature vectors, and two neural network baselines based on CNN and GRU.} \xian{A detailed description about our model architecture can be found in Fig.~\ref{fig:dpmodel} of Appendix~\ref{sec:dpmodel}. Besides, we present experimental settings such as hyperparameters and training details in Appendix~\ref{sec:dpparams}.}

\subsubsection{4.3.2 Performance and Interpretability.}
\begin{figure}[t]
    \centering
    \includegraphics[width=\linewidth]{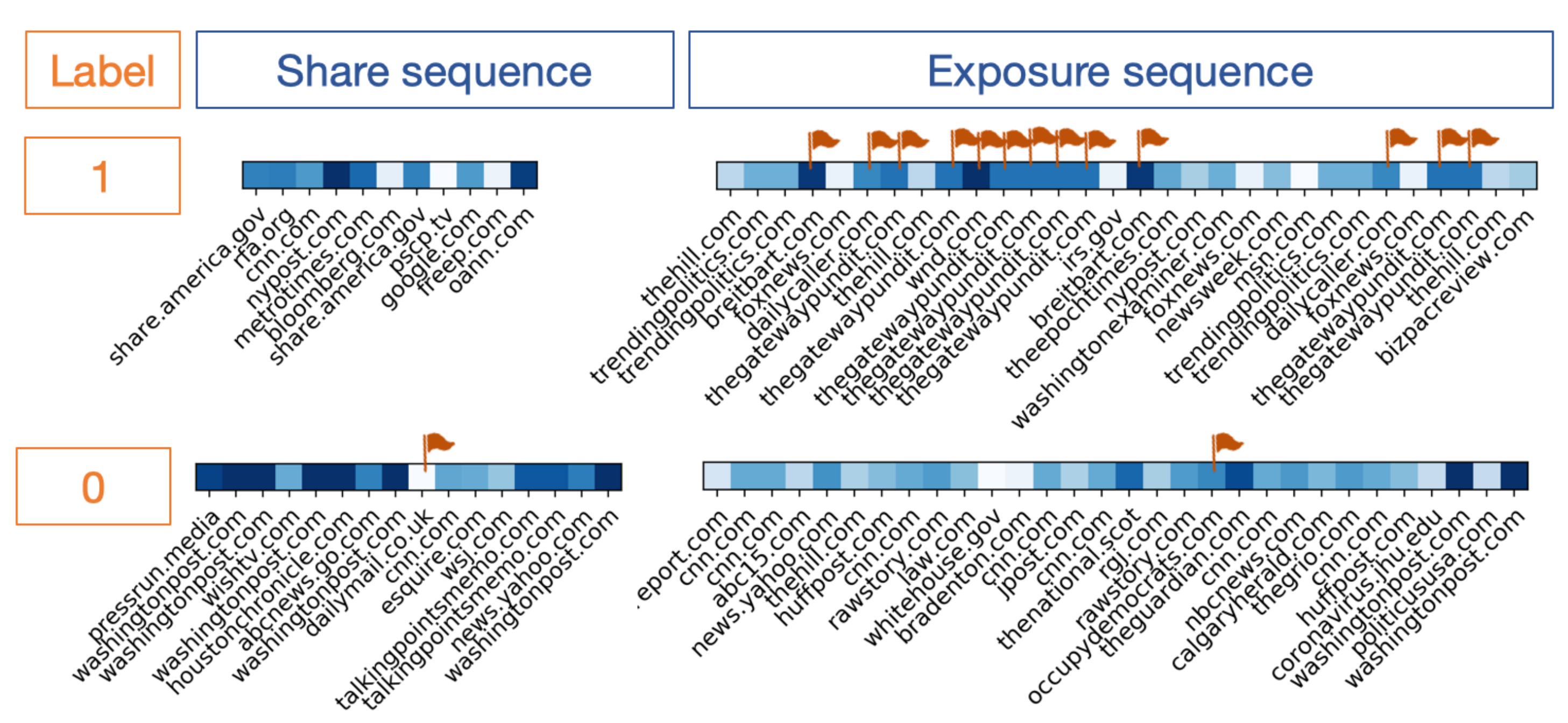}
    \caption{Contribution heatmap of domains from past share/exposure sequences. Darker color indicates larger contribution, flagged sources are marked in red.}
    \label{fig:cases}
\end{figure}

\xian{Table~\ref{tab:prediction} reports prediction performance: the overall result suggests that we are able to make reasonable real-time predictions simply based on users' short-term traces; besides, our model and other neural network baselines outperform the simple LR classifier. Fig.~\ref{fig:cases} shows two case study to illustrate the interpretability of our model, as it calculates a contribution weight to each of the past shared/exposed domains (darker colors indicates larger weights, red flags indicate if the news sources are flagged or not). In the first case, no past problematic shares were observed, but this user was heavily exposed to flagged sources over the past 2 weeks, our model allocated large attention on flagged domains from exposure (e.g., breitbart.com, thegatewaypundit.com) and on an unflagged, far-right, pro-Donald Trump cable channel (oann.com). One might suspect that the model purely generalizes past behaviors into the future. So we present the second case, both misinformation share and exposure were observed, but the model accurately predict a negative label and allocated large attention to mainstream or centered sources (washingtonpost.com, houstonchronicle.com, theguardian.com, talkingpointsmemo.com). We note that this model is unaware of any knowledge regarding a domain's learning or reliability, it learns domain-domain proximity through share/exposure sequences which in turn facilitates future prediction.}

\begin{table}[t]
\centering \fontsize{8}{9}\selectfont
\begin{tabular}{m{0.03\textwidth}m{0.085\textwidth}m{0.085\textwidth}m{0.085\textwidth}m{0.085\textwidth}}
& \makecell[c]{\bf accuracy} & \makecell[c]{\bf precision} &	\makecell[c]{\bf recall} &	\makecell[c]{\bf F1} \\
\midrule
LR&	0.696 (0.003)&	0.669 (0.004)&	0.761 (0.002)&	0.712 (0.003) \\
CNN&	0.719 (0.003)&	0.700 (0.015)&	0.759 (0.043)&	0.728 (0.013) \\
GRU&	0.719 (0.005)&	{\bf 0.701 (0.013)}&	0.755 (0.032)&	0.727 (0.010) \\
Ours&	{\bf 0.720 (0.004)}&	0.689 (0.007)&	{\bf 0.793 (0.012)}&	{\bf 0.737 (0.004)} \\
\end{tabular}
\caption{Model performance with average and standard deviation obtained from five trials of experiments.}
\label{tab:prediction}
\end{table}

\subsubsection{4.3.3 Dynamics of Misinformation Ecosystem.} 
\begin{figure}[t]
    \centering
    \includegraphics[width=\linewidth]{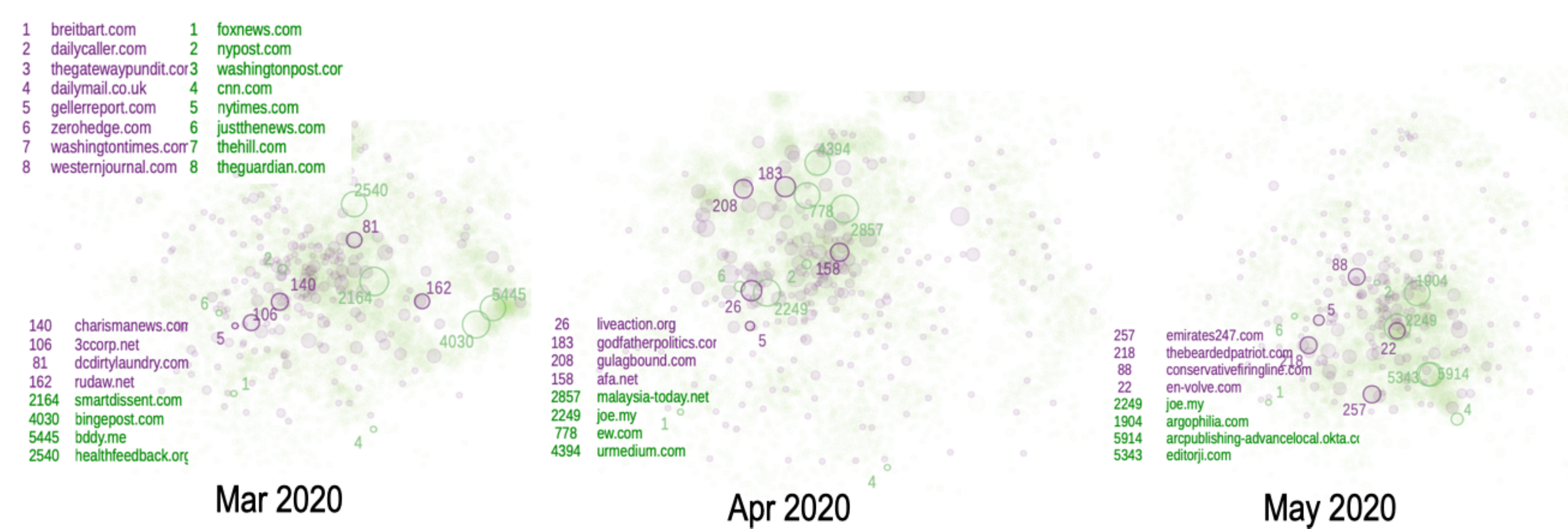}
    \caption{Domain proximity overtime. Nodes represent domains with color indicating domain type (purple: flagged; green: unflagged). Nodes are placed based on their pairwise proximities (derived from embeddings) and simultaneously retained at similar locations across weeks. Nodes were placed on approximately same location across weeks. Top list shows the most connected domains with the highest frequencies to be shared over the three months, bottom lists show the most connected domains emerged from the corresponding week. \xian{A high resolution version of this figure is provided in supporting file.}}
    \label{fig:news_ecosystem}
\end{figure}

\xian{Besides the temporal variation of individuals' susceptibility, we also examined the dynamics of misinformation ecosystem through a domain proximity network. Fig.~\ref{fig:news_ecosystem} shows the networks for three different weeks, the first week of March, April, and May. Nodes are placed based on their proximities with every other node and simultaneously retained at similar locations across weeks. For the clarity of presentation, only 50\% of nodes 624 (over 6K nodes) with the highest weighted degrees are shown. Comparing the snapshot networks, we observe that, while several most connected nodes seemed to reappear within similar clusters, there are new nodes emerging as new cluster centers as well. To see this, we list, on the top, the top 8 domains with the highest frequency appearing in users’ sharing history, and also list the most connected domains having the highest degrees at the bottom for each network of the week. We found the bottom lists change rapidly, suggesting that, week by week, different flagged sites become dominant in the network as they emerged from the sequences of users’ sharing. We further compute the correlation of the (weighted) degree of nodes across time, and found only 13-19\% correlation (in terms of Spearman’s rank correlation) between any two consecutive weeks. This result highlights the changing nature of users’ sharing behavior, but at the same time demonstrate the feasibility of utilizing a deep learning framework to capture the emerging flagged domains that put new risk of misinformation to users.}

\section{Discussion}

This study estimated the extent to which different user types contributed to COVID-19 misinformation spread. Our study suggests that around 90\%-99\% accounts that spread misinformation were humans instead of bots, over 50\% bots were benign. \revision{This finding partially aligns with recent work \cite{gonzalez2021bots} indicating that bots are less central than verified accounts during contentious political events, and our work provides new evidence of bot prevalence in COVID-related misinformation consumption.} Regarding humans' heterogeneous misinformation behaviors, strong-adherent users exhibited more aggressive sharing activities and seemed to be more successful in spreading compared to their weak-adherent counterparts. The difference might be unsurprising as we anticipate users might have different intentions -- for example, strong-adherent users might aim at supplying or/and promoting misinformation for certain purposes, whereas weak-adherent individuals might be occasionally exposed to misinformation but easily triggered to further spread misinformation.
We note that, though previous work have shown concentration of volume in misinformation consumption \cite{grinberg2019fake,guess2018selective}, little efforts have been made to fully understand the difference in intentions of susceptible users, which might lead to the difference in sharing tactics, temporal dedication and spread influence.

\revision{Based on our findings of different misinformation behaviors among strong-/weak-adherent users, platforms might consider devising different interventions tailored to different subgroups. For example, platforms could adopt disincentives policies \rrv{\cite{grinberg2019fake}} to reduce the visibility of content from strong-adherent users (i.e., misinformation suppliers); on the other hand, platform could proactively deliver fact-checking corrections to weak-adherent users through their congruent friendships to boost resilience against misinformation, as our results have shown that average users have a stronger tendency to be influenced from friends, prior works also demonstrated political neutral users likely appreciate corrections of fact particularly when issued through friendships. \cite{parekh2020comparing,hannak2014get,margolin2018political}. \rrv{Future work should examine} the effectiveness of such tailored intervention strategies in limiting misinformation cascades on the platform.}



\addr{Our observation about political affiliation is consistent with prior work \cite{grinberg2019fake}, which is expected due to the domain list we partially relied on -- more political news sites among the list, and more right-leaning sites among the lower-credibility list. However, instead of focusing on political news, our study focuses on the COVID-related topics. Moreover, we found that, while CSUs are more likely than CRUs to be associated with right-leaning exposures, they tend to interact with both supporting and opposing party leaders (though in distinctive ways) as well as use both left/right-leaning hashtags in profiles. Our results suggest that, even with this global health issue, CSUs may be more responsive to politically charged COVID-related content and interactions. Our observations about emotional reactions deviates from the prior study on rumors \cite{vosoughi2018spread}, where no substantial differences were observed in users' negative emotional reactions when interacting with false (vs. true) news. Unlike the spread of misinformation in normal times, our result suggests that during this COVID crisis, susceptible users may be more responsive to negative-emotional charged problematic content. The emotional responses were suggested by many crisis/risk literature \cite{inbar2009conservatives} but we present the first large-scale empirical observation. Finally, our observation about the relationship to pre-existing susceptibility suggests the importance of looking at user susceptibility as a dynamic and situational status rather than a intrinsic trait. The proposed model could be deployed in settings where falsehood or rumors need to be paid attention early before they become entrenched in public debate, particularly during uncertain and critical times like this pandemic.}

Our study has limitations. The user panel was obtained from snowball sampling and through a series of criteria, which might not be able to reflect the demographics of US population, the registered accounts on Twitter or other platforms. Another shortcoming comes from the definition of misinformation at news source level instead of at story level, along with the construction of source dictionary. \revision{Our study didn't account for the misinformation that existed outside of articles/URLs, such as false claims in text, images or \rrv{videos, nor did we consider the cases} that flagged sources might publish a mixture of misinformation and true stories.} Besides, the major presence of conservative news sources in flagged source dictionary might partially explain the association between conservative leaning and misinformation consumption. \revision{Our findings of bot prevalence is limited by the bot detection tool. Botometer retrieved the most recent tweets to assess bot-like behaviors, the scores might be partially based on tweets posted before/after our study period. 
Bots are continuously evolving and novel behaviors emerge, thus our findings to some extent depend on the tool's capability in capturing such novel patterns.} In addition, our analyses and prediction of linking online behavioral features to the tendency of sharing misinformation are purely correlational, more work are needed to explain why certain subgroups of people tend to share misinformation in this pandemic. Finally, though deep learning models achieved promising performance in our prediction task, they were not immune to pre-existing biases in training data, therefore they might be trained to better capture patterns from dominant subgroup of users.

\section*{Acknowledgement}
The authors would like to acknowledge the support from NSF \#1739413, \#2027713, and AFOSR awards. Any opinions, findings, and conclusions or recommendations expressed in this material do not necessarily reflect the views of the funding sources.

{\fontsize{9}{9}\selectfont \bibliography{reference.bib}}

\section{Appendix}
\subsection{Deep Learning Model}\label{sec:dpmodel}

\begin{figure}[ht]
\centering
\includegraphics[width=0.8\linewidth]{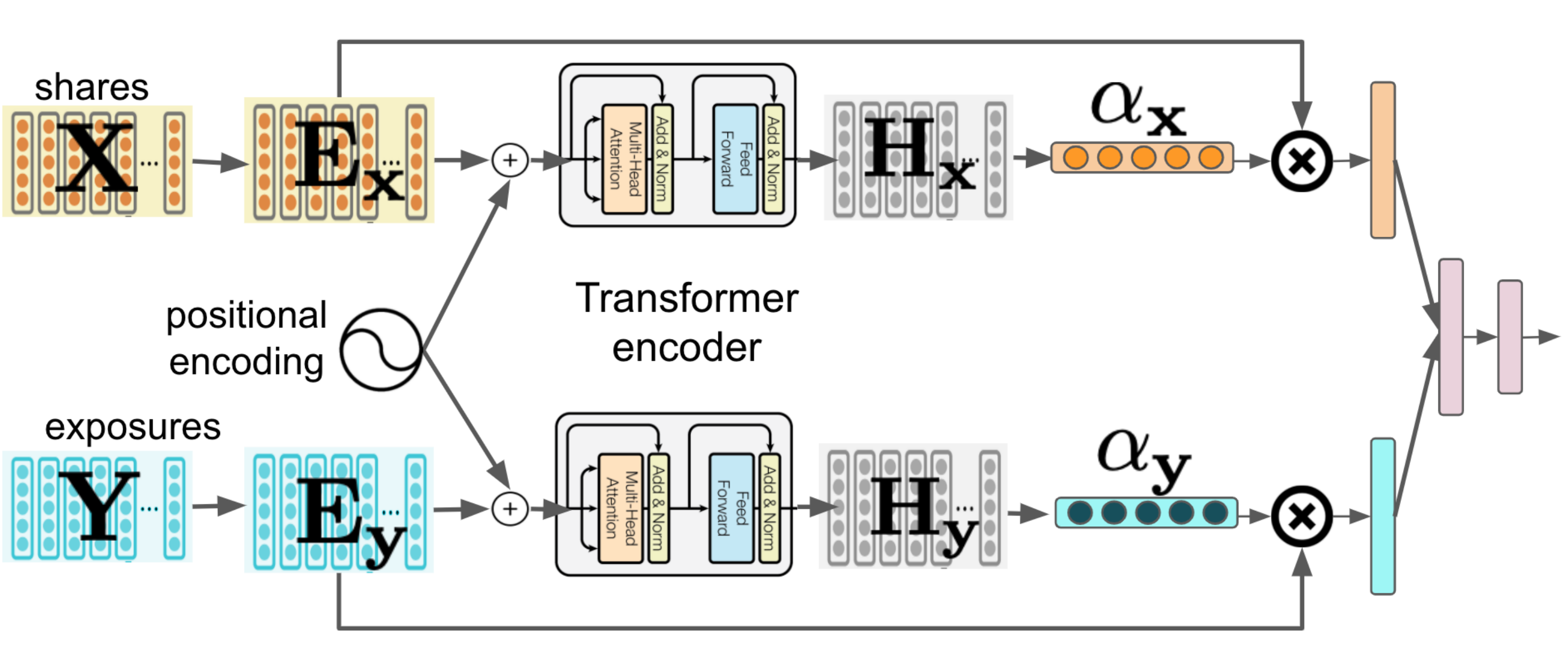}
\caption{Deep learning model.}
\label{fig:dpmodel}
\end{figure}

The model was inspired by prior work \cite{zhang2020inprem, vaswani2017attention}. In Fig.~\ref{fig:dpmodel}, both inputs -- the share sequence $\mathbf{X}$ and the exposure sequence $\mathbf{Y}$ -- consist of one-hot columns which could be mapped into embeddings $\mathbf{E}_{\mathbf{x}}$ and $\mathbf{E}_{\mathbf{y}}$, respectively. These embeddings are further encoded by a Transformer encoder into corresponding hidden states $\mathbf{H}_{\mathbf{x}}$ and $\mathbf{H}_{\mathbf{y}}$. The hidden states are in turn used to obtain two coefficient vectors $\alpha_{\mathbf{x}}$ and $\alpha_{\mathbf{y}}$ to indicate the contribution weights of domains from share and exposure sequences. We multiply domain embeddings with the corresponding coefficients to obtain a final representation for each input channel. Finally, a final output layer takes the concatenation of two vectors as input and outputs a probability $\hat{y}$ to indicate the probability of future susceptibility.

As shown in Fig.~\ref{fig:dpmodel}, the output probability $\hat{y}$ can be described as below:
{\fontsize{9}{6}\selectfont
\begin{equation}
\begin{split}
    \hat{y} &=
    \mathrm{Softmax}(\mathbf{W}_{o}[\mathbf{h}_{\mathbf{x}},\mathbf{h}_{\mathbf{y}}] + \mathbf{b}_{o}) \\
    &=\mathrm{Softmax}(\mathbf{W}_{o}[\alpha_{\mathbf{x}}\odot\mathbf{W}_{emb}\mathbf{X},\alpha_{\mathbf{y}}\odot\mathbf{W}_{emb}\mathbf{Y}] + \mathbf{b}_{o}),
\end{split}
\end{equation}} where $\mathbf{W}_{emb}$ indicates embedding parameter and $\mathbf{W}_{o}$ ($\mathbf{b}_{o}$ is the weight (bias) of output layer, $\odot$ means element-wise multiplication, $[\cdot,\cdot]$ indicates the concatenation operation of two vectors. Accordingly, the contribution of domain $i$ at time $t$ to the final prediction is
{\fontsize{9}{6}\selectfont
\begin{equation}
\begin{split}
    CB[t,i] &=\alpha_{\mathbf{x}}[t]\cdot\mathbf{W}_{o}^{\mathbf{x}}\mathbf{W}_{emb}[:,i],
\end{split}
\end{equation}}
\noindent where $\mathbf{W}_{emb}[,i]$ is domain $i$'s embedding, and $\mathbf{W}_{o}^{\mathbf{x}}$ is the weight from $\mathbf{W}_{o}$ corresponding to input channel $\mathbf{X}$.

\subsection{Experimental Settings}\label{sec:dpparams}
\subsubsection {\bf 6.2.1 Baselines \& Parameters.} For our model, the feedforward dimension is 1024, the number of heads is 4, the number of encoder layers is 2. For GRU, we used bi-directional module to encode information from both channels and then concatenated the outputs for the final fully connected layer. The CNN has a combination of multiple convolution layers (with distinct filter sizes 2, 3, 4, 5) and max pooling layers. For all three neural networks, the dropout is 0.8 and the embedding dimension is 64. For the logistic regression classifier, we used the optimization solver ``lbfgs.''

\subsubsection {\bf 6.2.2 Training Details.} \revision{We prepared training data based on COVID-19 shares between March 1 and June 1, 2020. Since the panel produced a huge amount of data, we randomly sampled 30K users and we generated share/exposure sequences (observed in a 2-week period) and output labels (obtained in a 1-week period) in a sliding window manner. Long sequences of shares/exposures were truncated at maximum length 50, short sequences were abandoned at minimum length 5. After this procedure, there are 17735 unique users left and about 73000 samples as our data. The percentage of positive samples is 49\%. In the training procedure, we shuffled and randomly split the samples into train 60\%, validation 20\% and test 20\%. We ran the training procedures for 5 times and reported the average values and standard deviations for evaluation metrics.}
\end{document}